# A Data Transparency Framework for Mobile Applications

Steven C. Isley, Adjunct Staff Member, *The RAND Corporation*

*Abstract*— **In today's mobile application marketplace, the ability of consumers to make informed choices regarding their privacy is extremely limited. Consumers largely rely on privacy policies and app permission mechanisms, but these do an inadequate job of conveying how information will be collected, used, stored, and shared. Mobile application developers go largely unrewarded for making apps more privacy conscious as it is difficult to communicate these features to consumers while they are searching for a new app. This paper provides an overview of a framework designed to help consumers make informed choices, and an incentive mechanism to encourage app developers to implement it. This framework includes machine readable privacy policies encouraged by mobile app stores and enhanced by user software agents. Such a framework would provide the foundation required for more advanced forms of privacy management to develop.**

## I. INTRODUCTION

PRIVACY is increasingly a concern in today's digitally connected world. Information about people, much of it private, is being collected and stored in an ever increasing number of ways and places. An individual's relationships used to be known by a small group; now that same information is stored on their phone, by their cellular service provider, multiple social networks, email providers, gaming platforms, and an array of mobile applications residing on their smartphone or tablet.

This last category, the apps on their smartphone or tablet, are the focus of this paper. Mobile devices are particularly concerning from a privacy point of view as they are carried with people on a daily basis and can be used to collect and transmit information at fine geographic and temporal resolutions. This enhanced data environment enables many new and useful functions but comes with a concomitant rise in the risk of privacy failures.

Mobile applications are part of and supported by an ecosystem of software, hardware and firms. What this ecosystem needs is a functional mechanism for ensuring that an individual's privacy orientation is respected. The system needs to work for users across the spectrum of technological sophistication in a way that is unobtrusive yet provides assurance that data about an individual is being used in a transparent way.

This paper will argue for a data transparency framework that includes machine readable privacy policies encouraged by mobile app stores and enhanced by user software agents. Each of these components will be discussed in turn.

Part one describes the mobile application privacy status quo and recent work on implementing machine readable privacy policies. Part two focuses on how mobile app stores could implement such programs (and indeed are in a unique position to do so). Part three explores how user agents can augment the services provided by app stores and help users make and maintain informed decisions.

## II. PRIVACY IN MOBILE APPLICATIONS

Before discussing mobile applications, it is worth defining what is meant by privacy. A general conception of privacy is hard to come by, with Daniel Solove on page one of his book, Understanding Privacy, going so far as to say that nobody can articulate what it means [1]. While this paper is not intended to argue the merits of privacy, I will limit my discussion of privacy to mean the rights of individuals to control the collection, use and dissemination of their personal information.

An interesting exercise is to think of personal information as a marketplace. This idea is certainly not novel, with Meglena Kuneva, the European Consumer Commissioner, calling data the new oil of the internet [2]. Much academic research exists on the personal data economy with issues such as who owns data [3], the economic downside of privacy [4], and issues of marketplace self-regulation [5].

An important component of any public marketplace is transparency. The Securities and Exchange Commission exists for this purpose. If insider trading were legal then an efficient public market could not exist as potential investors would realize their knowledge disadvantage relative to insiders. Rather, insider trading is illegal with often severe prison sentences for those found engaging in it. The mobile data economy suffers from a similar lack of transparency: data collection, use, and dissemination practices are largely 'insider' information and those that provide the data, the users, are often wary of revealing it (of participating in the market), like the potential investors from before. This hesitancy to reveal personal information could lead to lost economic gains. The withdrawal from the mobile information marketplace is not theoretical. According to the Pew Research Center, more than half of all smartphone owners have avoided an app due to privacy concerns and 30% have uninstalled an app after they learned it was collecting personal information they did not wish to share [6].

Given this information asymmetry in the information market, surprisingly few mechanisms have been developed to help individuals understand or control how their data will be used. These include privacy policies, app permission mechanisms, privacy certification services, app analysis tools and the occasional investigative report. Of these, I will focus on only the first two as they are widely used by companies (privacy policies) and the public (app permissions). However, for an excellent example of an app analysis tool, see the AppBrain Ad Detector [7] and for investigative journalism, see the Wall Street Journal's "What they Know Mobile" series [8].

*A. Privacy Policies*

A privacy policy is typically a special page associated with an app, website, or firm that describes how data are collected, used, or shared. Often it includes information about how to opt-out of such practices. Such polices proliferated rapidly at the turn of the century, going from 2% of all websites in 1998 to almost all popular websites by 2001 [9]. Ideally, these policies would convey to readers what data are collected, how the data are used and who the data are shared with. However, in practice, privacy policies exist more to protect firms from litigation than they do to inform users, and they can be error-prone and difficult to read [10, 11]. This stems from how the FTC has prosecuted firms for violating their own privacy policy as a deceptive or unfair business practice. The resulting incentive is to make a bland and generic privacy policy with vague mentions of sharing information with 'trusted business partners' or collecting 'other usage data' that is written in legal language that is particularly difficult for the laymen to understand.

Recognizing these shortcomings, the W3C attempted to implement a technology called the "Platform for Privacy Preferences" or P3P. This standard specified a structured document that would describe how an organization treated data. The structure allows a computer to read and understand the document (it is 'machine readable') and then parse it into natural language. Such software is called a P3P user agent, and it acts as an intermediary between data collectors and users. It can be configured to aid the user in avoiding websites that have data practices with which they are uncomfortable. However, the standard lacks any enforcement mechanism since no real connection exists between the P3P policy and the firm's actual usage of personal information.

P3P has many other shortcomings as documented by Hogben [12] that include the lack of formal semantics and poor adoption amongst browsers. Mozilla's Firefox browser removed all P3P related code in the mid 2000's, and this author could find no browser extension compatible with the latest Firefox release. While P3P is not at present a viable solution, it may provide a building block for better privacy languages. Work in this direction has already been completed [13].

In the mobile apps ecosystem, privacy policies exist for the majority of apps. Links to these are provided in the Google and Apple app stores. However, these privacy policies suffer from all the drawbacks mentioned previously; they are not machine readable, they are written in legal language and they are overly vague. For example, the Angry Birds mobile app (owned by Rovio) has the following statement in its privacy policy:

"Rovio or third parties may collect and use data, for [advertising] purposes, including but not limited to, data such as IP address, …, unique identifiers in browser cookies,…, Internet and on-line usage information and in-game information" and,

"Rovio retains the collected data for the period necessary to fulfill the purposes outlined in this Privacy Policy unless a longer retention period is required or permitted by law. Thereafter Rovio deletes all aforementioned data in its possession within a reasonable timeframe." [14]

These statements are extraordinarily vague, but typical of privacy policies, noting that other data may be collected and that it can be retained as long as Rovio wants or is legally allowed. It is also unclear which portions of the privacy policy apply to Angry Birds as opposed to any of Rovio's websites. Nowhere are the referenced 3rd parties actually named, so even if a user wanted to look at these other companies' privacy policies they would have no ready way of doing so. Some independent analyses have been completed. Jason Hong of Carnegie-Mellon identified eight different ad networks included in the free version of Angry Birds [15].

After reading a few privacy policies, one can easily see how time consuming it would be to read through, analyze, and incorporate the information from them into a decision concerning which of 30 exercise apps to choose, or 25 personal finance apps.

*B. App Permissions*

Both Android and iOS include permission mechanisms to limit access to personal information from mobile apps. Android uses a system of install-time warnings where users are presented with a list of required permissions and must accept them or forgo installing the app. Apple's iOS uses a combination of runtime confirmation dialogs and review screens. The runtime confirmation dialogues appear the first time an app requests a particular permission, like access to location information, and users can deny the request and continue using the app. A review screen is displayed by iOS before certain actions, like sending a text message, allowing the user to see which application is trying to send it and giving them the opportunity to edit or dismiss the text.

Both systems suffer from a lack of specificity in their permission models. No indication of why an app needs a particular permission is provided and Android's install-time warnings are often ignored or misunderstood by users [16]. This lack of specificity manifests itself in two ways. First, some apps ask for permission they legitimately need but do not seem to require at first glance. E.g., a calendar app may request access to contact information in order to better arrange meetings. Second, some apps request two separate permissions with obvious need, but use them together for non-obvious purposes. E.g., a traffic navigation app could request both internet access and location information (for obvious reasons) but then use those permissions to send location information to advertisers (a non-obvious purpose).

*C. What is Needed*

In order to provide real transparency, users need to have access to the types of information collected by an app and the context in which it is collected, used, stored and shared. A data's context is the full picture surrounding it. It's not enough to know that an app has access to location information; in fact, that tells one very little. An app might have access to location data, but only use it to provide a specific service and never record or transmit it. The privacy risk in this situation is negligible. Rather than simply knowing what data an app might have access to (which is all a permission conveys), users need to know the context surrounding their data. Users should be able to easily find out what their data are used for, how often it is collected, where it is stored, how long it will be stored, how the data are connected with them (e.g. by their name, a device identifier, a hashed version of either), what sort of security protections are in place (when stored) and who it can potentially be shared with and for what purpose. These additional details, the context, provide the foundation upon which informed choices can be made.

This sort of information should be readily available in privacy policies, but it is not. Even if it were included, users would face an insurmountable hurdle gathering the information for the subset of apps they are considering and incorporating it into their decision. Nor can that information be used to decide on the subset of apps to consider in the first place, which would require gathering information for all apps of possible interest. In the end, apps that do a better than average job of respecting a user's privacy aren't rewarded.

The way to change this would be to place all that relevant data use information in a machine readable privacy policy (MRPP). This could be based on the previously mentioned P3P framework but would need to be made more precise. Such a document would contain a description of the collected data and their contexts; and, importantly, links to the MRPP (or domain name if none exists) of any 3rd parties that they share data with and those they receive data from. By requiring apps to list both data coming in and data leaving, a layer of accountability is provided. Two groups must omit the relationship in order to hide the transfer of data. It also provides an audit mechanism. Researchers or the app store itself can easily find one-way references and work to clarify the situation.

With enough participation, the full implications of sharing personal information with a particular app can finally be realized. Each app would generate a web of data sharing going out potentially many degrees and the size and composition of this web could be used to help inform user choices. While analyzing a web of data is more tractable than a host of non-standardized documents full of legalese, it is still beyond the technical ability of most and the attention span of nearly all. The final section of this paper will discuss mechanisms for easing the burden on the end users, illustrated by potential use cases. But first the issue of getting 'enough participation' will be addressed by looking at the incentives and capabilities of app stores.

### III. THE PRIVACY AWARE APP STORE

App stores are uniquely situated to increase the transparency of mobile applications to end users. Unlike websites, there are relatively few app stores and this affords a degree of centralization that can be used to encourage increased transparency. However, is it in the interests of app stores and app developers to promote increased transparency? An analysis of the incentives faced by each may yield policies that can lay the groundwork for a transparency 'race to top.'

Currently, very few incentives exist to reward app developers for harvesting less personal information or being more transparent about the ways they use and share data. The incentives are sticks rather than carrots – an app can suffer negative publicity or reviews if users find out their data isn't handled properly. Both Google Play and the Apple App Store have rudimentary search features. This is particularly surprising in the case of Google as they have more combined experience in search than any organization in human history. Currently, it is not possible to do even simple searches like "games that don't require internet access" in either app store. Privacy conscious users could greatly benefit from being able to see search results tailored to their preferences.

However, even if that capability were added, it would still be insufficient for the aforementioned reasons – the ability to access certain data does not determine how it is actually used or shared. For that type of transparency a MRPP is needed. An app store could require this of all apps, but that might engender significant push back from app developers due to the burden of creating and maintaining another document using a new syntax. Some developers will certainly be wary about revealing their data collection and use practices. This could reduce the number of high quality apps in the store, which is one of the main metrics stores use to compete. Aside from simply requiring that all apps create a MRPP, there are many incentives that app stores could put in place that could encourage large numbers of app developers to participate. These incentives fall into two categories, improving app discovery and app adoption.

The problem of app discovery stems from the fact that there are now more than one million apps on Google Play [17], and over 900,000 on Apple's App Store [18]. As an app developer, it is difficult to get your app noticed amongst the millions of competing apps. Specialized services have sprung up to help with this specific problem [19]. App developers will often pay advertising networks to include links to their app in banner ads shown on mobile devices. Fisku is one such advertiser and they report that the average cost of acquiring a loyal user was

$1.90 in August of 2013 [20]. Anything that will make an app easier to find by app store users will serve as a strong incentive. App stores use some internal ranking algorithm to present search results to users, and this could be modified to boost the rank of apps that include a validated MRPP. Alternatively, users could be given the ability to search amongst only those apps that have a MRPP, or by specific characteristics that only a MRPP would reveal. App developers might be particularly interested in these types of searches as they might be utilized more by individuals willing to pay for apps that protect their privacy.

A second avenue to encourage the adoption of MRPP's would be an icon or seal placed on an app's download page that notified the viewer that the app was participating in the app store's voluntary transparency program. Various types of awards could be created to show the app's level of transparency or minimal data collection policy. For instance, a bronze award could be given to an app that has a MRPP, a gold award to apps whose first degree sharing partners all have MRPP's, and a platinum award for apps whose entire data sharing web consists only of members with a MRPP. This would provide an incentive for app developers to work with companies that employ MRPP's, even firms that don't produce apps. Data aggregators have emerged in recent years as a highly opaque group of firms that buy and sell personal and non-personal data via large electronic exchanges. These operators are far from view and face very few accountability constraints. By rewarding app developers for only sharing data with firms that publish a MRPP, a strong incentive would emerge for these aggregators to become transparent, or risk losing business to another aggregator that does.

An app store wishing to promote MRPP's must select a simple set of semantics and provide a set of tools that makes creating the policy simple. Ideally, this would start by analyzing an app's existing set of permissions and creating a framework on which the additional information could be added easily. Additional tools could analyze a developer's draft MRPP and highlight what is needed to get the various icons promoted by the app store.

With a foundation of MRPP's, an app store could even go one step further by tailoring their customers' search results to conform to their customers' stated privacy preferences. Individual apps could be given a personalized star rating that would be unique to each customer. This concept overlaps with the notion of privacy user agents, the subject of the next section.

## IV. Privacy User Agents

In a world of MRPP enabled apps, imagine if an unknown, behind the scenes data aggregator had a data breach. Individuals are signed up for one of many services that monitors such things. These middle agents then update all their users' policies to reflect that this particular actor is blacklisted. Now anytime those users visit a website or use a mobile app that has a connection with that data provider (regardless of how far away that connection is) they will receive a notice informing them of the new threat to their privacy. They can choose to ignore it, but some users will see this and choose to visit a different website or use a different app. The resulting drop in membership will cause some of those websites and apps to sever their relationships with that particular data aggregator, and this will lead to real economic losses to the aggregator that didn't make security a priority. What was before an opaque relationship with accountability far removed from actions suddenly becomes a transparent relationship with immediate consequences. No government regulations were needed for this to occur, just well positioned actors operating (and potentially profiting) in the new privacy space.

With MRPP's in common use, a host of new tools and services would evolve to take advantage of the information. User agent apps would solicit the privacy preferences of their owner and use those to monitor the privacy policies of the other apps installed on the system. If an app's MRPP is updated in a way that conflicts with the user's privacy settings an alert is shown. It could even recommend other replacement apps and generate revenue in this way. Such apps could even be included with the OS by default, a sort of 'stock' privacy watchdog that can be replaced with a more feature rich one by a user at their discretion. These watchdog apps could communicate with one or more app stores to provide the sort of personalized recommendations described in the prior section. A privacy ecosystem could evolve to complement the data ecosystem and provide users with the tools they need to navigate this new, complex space.

Another aspect of having MRPP's that may be somewhat controversial is their use in law enforcement. As more communication moves into the digital world, traditional methods of gathering evidence for investigations become less useful. Law enforcement has a legitimate need to investigate crimes, and this means accessing personal information given appropriate legal oversight. A largely ignored overlap between law enforcement and privacy advocates concerns data transparency: both groups want to know who has what information, how long they keep it and who they share it with. It is hard to argue this information should be made available to the public but somehow kept from law enforcement. A system of MRPP's would help police investigators know where data may reside and if resources should be spent requesting it through existing legal channels.

## V. Conclusion

Machine readable privacy policies could provide the transparency that privacy conscious users need in order to make informed decisions. They could provide economic benefits for those app developers willing to cater to this market, and app stores are in a unique position to provide the incentives necessary to promote adoption. The resulting web of linked MRPP's would provide a sorely needed accountability mechanism that is needed to protect individual privacy and fully exploit the potential gains from users sharing their personal information with mobile applications and other 3rd parties.

The first major app store to sufficiently encourage this type of transparency could receive an image as the app store that truly values its customers' privacy. Less popular mobile operating systems and their accompanying app store could differentiate themselves from their more successful competitors along a customer preference dimension that has yet to be catered to. It could provide the data for researchers to show the extent of the mobile data economy and ways it can be made more secure while simultaneously providing more benefits for consumers. This could lead other app stores to implement similar policies in order to remain competitive.

This is not the be all and end all of solutions, but a workable first step that could foster the creation of more services geared towards privacy conscious users. Just as creating the app distribution framework led to an explosion of apps with features never predicted, providing a framework for privacy analysis could lead to privacy enhancements we can't predict. However, these won't emerge unless the underlying resources are available.

The mobile app ecosystem is a highly competitive, dynamic market place. It's time that privacy and transparency became central actors in this sphere, with influence, incentives and accountability mechanisms that operate at times against, at times with, the current set of interests.

ACKNOWLEDGMENT

The author would like to thank Ed Balkovich of the RAND Corporation for invaluable feedback on earlier drafts of this manuscript.